\documentclass[reprint,amsmath,amssymb,prb,floatfix]{revtex4-2}

\usepackage{hyperref}
\usepackage{graphicx}
\usepackage{bm}

\usepackage{xcolor}

\begin{document}

\title{Low-energy in-gap states of vortices in superconductor-semiconductor heterostructures}

\author{Alexander Ziesen}
 \email{alexander.ziesen@rwth-aachen.de}
\author{Fabian Hassler}
\affiliation{JARA Institute for Quantum Information, RWTH Aachen University, Aachen, Germany}

\date{March 2021}

\begin{abstract}
The recent interest in the low-energy states in vortices of semiconductor-superconductor heterostructures are mainly fueled by the prospects of using Majorana zero modes for quantum computation. The knowledge of low-lying states in the vortex core is essential as they pose a limitation on the topological computation with these states. Recently, the low-energy spectra of clean heterostructures, for superconducting-pairing profiles that vary slowly on the scale of the Fermi wavelength of the semiconductor, have been analytically calculated. In this work, we formulate an alternative method based on perturbation theory to obtain concise analytical formulas to predict the low-energy states including explicit magnetic-field and gap profiles. We provide results for both a topological insulator (with a linear spectrum) as well as for a conventional electron gas (with a quadratic spectrum). We discuss the spectra for a wide range of parameters, including both the size of the vortex and the chemical potential of the semiconductor, and thereby provide a tool to guide future experimental efforts. We compare these findings to numerical results. 
\end{abstract}

\maketitle

\section{\label{sec:intro}Introduction}

The research of in-gap states of vortices started in the 1960s when approximate solutions to the Bogoliubov–de Gennes (BdG) equations for a vortex in pure type-II $s$-wave superconductors (SC-s) were found and the scaling of the states bound to the vortex core was determined as $\Delta_{\text{SC}}^2/E_F$; with $\Delta_{\text{SC}}$ the bulk superconducting gap and $E_F$ the Fermi energy \cite{CAROLI-1964,Bardeen1969}. Those states arise due to the phase winding of the superconducting order parameter around the normal core \cite{Berthod2005} and their scaling was verified numerically in Ref.~\cite{Schlueter1991}. The study of in-gap states became even more exciting when it was realized that vortices of chiral $p$-wave superconductors exhibit zero-energy Majorana modes \cite{Volovik1999} that braid as non-Abelian particles \cite{Read1999,Ivanov2000}. As there are no superconductors known that unambiguously show a chiral $p$-wave pairing, these ideas remained academic for some time.

Recently, it was realized that the combination of a conventional $s$-wave superconductor with the surface states of a topological insulator (TI) in a heterostructure emulates the same physics \cite{FuKane2008}. 
An $s$-wave superconductor is deposited on the surface of a three dimensional topological insulator gapping out its surface states except for a closed inner region that is left uncovered. Within this unproximitized region, a superconducting flux quantum is trapped which introduces the desired phase winding and confines a Majorana mode at zero energy. It was immediately realized that the design flexibilities of the heterostructure approach also allow to design the minigap, \emph{i.e.}, the gap of the Majorana mode to the first excitation \cite{Das_Sarma2010}. In fact, typical values of $\Delta_{\text{SC}}^2/E_F$ in natural superconductors are smaller than 1\,mK, rendering the level spectrum inside a vortex core akin to a continuum. For heterostructures, the Fermi energy $E_F$ of the superconductor is replaced by the much smaller value inside the semiconductor, thus allowing to reach values for the minigap in the order $1$\,K \cite{Das_Sarma2010}. As such spacings are resolvable by tunneling spectroscopy \cite{Esteve2008,Feng2016,Herrero2017}, this insight spurred a renewed interest in understanding the in-gap spectrum of heterostructures in details \cite{Das_Sarma2010,Feigelmann2012,Nori_2014}.

For superconducting-pairing profiles varying slowly on the scale of the Fermi wavelength of the semiconductor, Refs. \cite{deng2020bound} analytically calculated the low-energy spectrum of heterostructures. They obtained the spacing of the states to be proportional to $\Delta_{\infty}^2/\mu$, where $\Delta_{\infty}$ denotes the bulk value of the superconducting gap in the proximitized semiconductor and $\mu$ is the chemical potential measured with respect to the charge neutrality point in the semiconductor.  

In this work, we employ an alternative method based on conventional perturbation theory  that allows to calculate the energy of the low-lying Andreev states of vortices given the pairing-potential and the magnetic-field profile as input. Our results are valid for semiconductor-superconductor heterostructures where the semiconductor is either a TI or a conventional two-dimensional electron gas (2DEG) with quadratic spectrum. The formulas allow to determine the effects of the many design parameters. Different from a natural vortex, in a heterostructure with an etched superconductor, the size of the vortex, the effective coherence length, and the chemical potential are all individually tunable. 

\begin{figure}[t]
  \centering
\includegraphics[width = \linewidth]{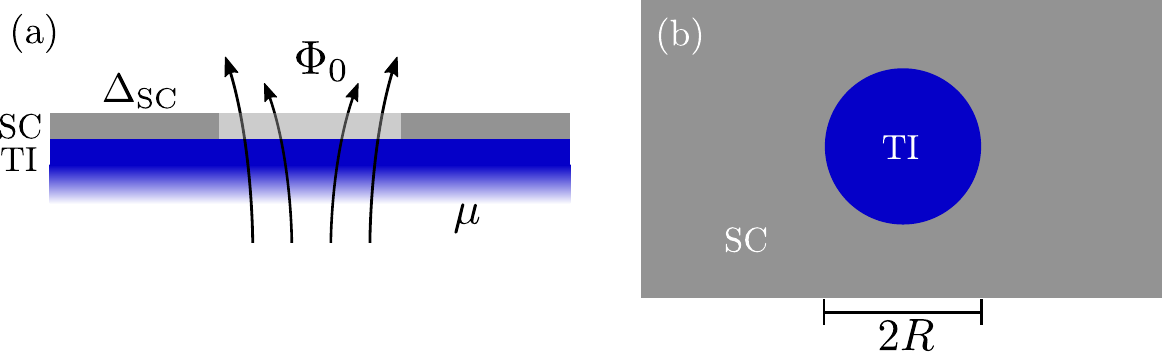}
\caption{Schematics of the topological insulator-superconductor heterostructure. Panel (a) is the side view, panel (b) the top view. The superconductor (gray) with pairing potential $\Delta_{\text{SC}}$ is deposited on the surface of a 3D topological insulator (blue) with chemical potential $\mu$. A superconducting flux quantum $\Phi_0$ is pinned to a circular region (light gray) with reduced pairing potential.}
\label{fig:syst_sketch}
\end{figure}

The paper is organized as follows. In Sec.~\ref{sec:phys_model}, we introduce the model for the SC-TI system. In Sec.~\ref{sec:energie_gaps_TI}, we provide results for the minigap in terms of the different design parameters. In Sec.~\ref{sec:de-Gen-syst}, the SC-2DEG heterostructure is introduced with the results for the minigap in Sec.~\ref{sec:de-Gen-spec}. Section~\ref{sec:osc_both} discusses oscillatory corrections in the vortex spectra in the case of a sharp gap profile. We conclude in Sec.~\ref{sec:concl}. Details of the numerical calculations performed in this paper are given in appendix~\ref{app:shooting}.

\section{The Topological Insulator heterostructure \label{sec:phys_model}}

Figure~\ref{fig:syst_sketch} displays a schematic of the heterostructure under consideration, see also Refs.~\cite{Das_Sarma2010,Nori_2014,deng2020bound} for related work.
An $s$-wave superconductor with a bulk pairing strength $\Delta_{\text{SC}}$ is deposited on the surface of a 3D topological insulator with chemical potential $\mu$ and a superconducting flux quantum $\Phi_0= \pi \hbar c/e$ is trapped on the surface where $e>0$ denotes the elementary charge. The closeness of the two materials results in a proximity induced pairing potential $\Delta_{\text{TI}}(\bm{r})$ on the surface of the TI \cite{Das_Sarma2010,Nori_2014}.
Given the bulk value $\Delta_{\text{SC}}$ in the superconductor, the exact function $\Delta_{\text{TI}}(\bm{r})$ is dependent on the transparency between the two materials \cite{Das_Sarma2010}. There, it has been found that for good transparency the $\Delta_{\text{TI}}(\bm{r})$ is proportional to $\Delta_{\text{SC}}(\bm{r})$. We drop the subscript TI in the following, since we always refer to the proximitized pairing function. We write the Hamiltonian of the proximitized TI surface in the BdG formalism as $H = \frac{1}{2} \int \!d^2r\, {\Psi}^{\dagger}(\bm{r}) H_{\text{BdG}}(\bm{r}) {\Psi}(\bm{r})$, with
\begin{equation} \label{eq:matrix_ham_bdg}
H_{\text{BdG}} = \begin{pmatrix}
	h_{\text{TI}}(\bm{A})-\mu & \Delta(\bm{r}) \\
	\Delta^{*}(\bm{r}) & \mu-h_{\text{TI}}(-\bm{A})
\end{pmatrix}\,
\end{equation}
and ${\Psi}(\bm{r}) = (\psi_{\uparrow}(\bm{r}), \psi_{\downarrow}(\bm{r}), \psi_{\downarrow}^{\dagger}(\bm{r}), -\psi_{\uparrow}^{\dagger}(\bm{r}) )^T
$ with $\psi_{\sigma}(\bm r)$ the fermionic field operators. 

The vortex at position $\bm r =0$ introduces a phase winding with $\Delta(\bm{r})= \Delta(r) e^{-i \varphi}$, $\Delta(r)>0$. Note that we have assumed that the setup is rotationally symmetric for simplicity. The vortex traps a superconducting flux quantum $\Phi_0 = 2\pi \int_0^\infty\!dr\, r B(r)$ with $B(r)$ the $z$-component of the magnetic field at the radius $r$ \footnote{To generalize to arbitrary vorticity $\nu$ and account for a different number of superconducting flux quanta trapped on the surface, we can introduce a factor $\nu$ in the phase of $\Delta(\bm{r})$. Furthermore, the flux generating the magnetic field is then multiplied by $\nu$.}.   The vector potential $\bm{A}$ has to be in the London gauge \cite{deGennes_book} and we find $\bm{A}(\bm{r}) = (1/r)\int_{0}^r \!ds \,s B(s) \bm{e}_{\varphi} = A(r) \bm{e}_{\varphi}$. The Hamiltonian for the surface states of the topological insulator is given by
\begin{equation}
	h_{\text{TI}}(\bm{A}) = v_D\bm{\sigma} \cdot \left( \bm{p} + \frac{e}{c} \bm{A} \right)\, ,
\end{equation}
with $v_D$  the Dirac velocity and $\sigma_j$ the Pauli matrices in spin-space. 

\begin{figure}[tb]
  \centering
\includegraphics[width = \linewidth]{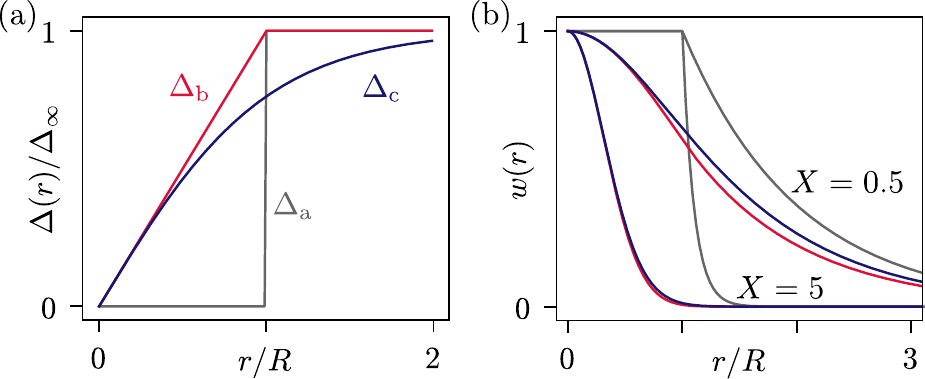}
\caption{Panel (a) displays the gap profiles $\Delta_x(r)$, we have considered. The profile $\Delta_\text{a}$ corresponds to a hole that is etched into the superconductor. On the other hand, $\Delta_\text{b}$ and $\Delta_\text{c}$ model a magnetic vortex where the pairing amplitude is suppressed due to the presence of the magnetic field and the resulting screening currents. All the profiles vanish for small $r/R$ and approach the bulk value for $r \gtrsim R$. Panel (b) shows the weighting functions $w(r)$, defined below Eq.~\eqref{averging}. Note that by decreasing $X=R/\xi$ the weight $w(r)$ is spread out to larger values of $r$.}
\label{fig:prof_sketches}
\end{figure}

\begin{figure}[tb]
  \centering
\includegraphics[width = \linewidth]{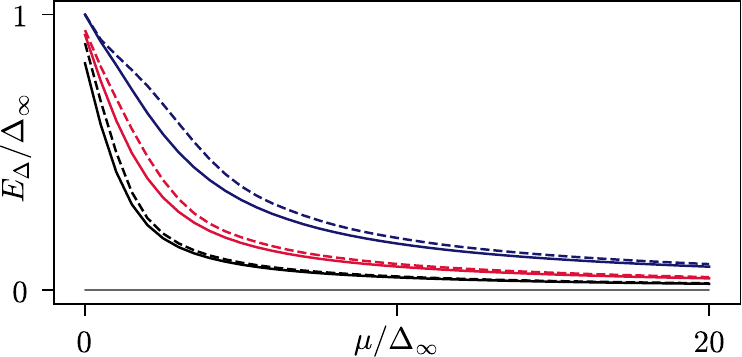}
\caption{Comparison of the numerically determined spectrum of the Hamiltonian $H_R$ as a function of the chemical potential $\mu$ for the pairing profile $\Delta_{\text{b}}(r)$ (dashed lines) with the one for $\Delta_{\text{c}}(r)$ (solid lines). The gray, red, blue lines depict the spectrum for the angular momenta $m \in \{0,1,2 \} $ at a radius $R=\xi$. The black line shows $m=1$ at $R=2\xi$. We see that the results for  $\Delta_{\text{b}}$ and $\Delta_{\text{c}}$ are the same with only small deviations at small $\mu$.}\label{fig:spec_comp_TI_ab}
\end{figure}

Due to the rotational symmetry, the Hamiltonian $H_\text{BdG}$ commutes with the total-angular momentum
\begin{equation}\label{Jz_TI}
	J_z = -i \partial_{\varphi} +\tfrac{1}{2}(\eta_z  + \sigma_z)\, 
\end{equation}
where $\eta_j$ are the Pauli matrices in particle-hole space. The states are thus classified by the eigenvalues $m$ of $J_z$ which leads to
the ansatz
\cite{DAS_vort_math_1983,Nori_2014,Marcus_full_shell}
\begin{equation}\label{trafo_angle}
  {\Psi}(\bm{r}) = e^{i (m -\frac{1}{2}\eta_z - \frac12 \sigma_z )\varphi
  -i \frac\pi4 \eta_z (1+\sigma_z) } \frac{F(r)}{\sqrt{r}} 
\end{equation}
with $F\colon [0,\infty) \to \mathbb{C}^4$. Note that the factor $r^{-1/2}$ ensures the Hermiticity of the radial equation below \footnote{We employ the additional phase factor $e^{  -i \frac\pi4 \eta_z (1+\sigma_z) }$ for convenience as it renders the radial equation real. }. Uniqueness of the wavefunction requires that $J_z -\tfrac12(\eta_z + \sigma_z) \in \mathbb{Z}$ which yields $m \in \mathbb{Z}$.
We split the radial Hamiltonian $H_R + H_A$ into two parts with the magnetic term $H_A = ( \pi \hbar v_D/ \Phi_0)A(r) \sigma_x\eta_z$  and the remainder
\begin{align}
  H_R =  \hbar v_D \biggl( i \sigma_y  \partial_r +\frac{2m- \eta_z}{2r} \sigma_x \biggr) - \Delta(r) \sigma_z \eta_x - \mu \eta_z \, .\label{radial_TI}
\end{align}
The eigenvalues $E$ of $H_R+H_A$ with $|E|<\Delta_{\infty}$ form the in-gap spectrum which is  
 particle-hole symmetric. This symmetry is implemented in the radial equation  by the mapping $(m,E)\mapsto(-m,-E)$. 
Note that $\Delta_{\infty}$ denotes the superconducting bulk gap away from the vortex, \emph{i.e.}, we assume
\begin{equation}\label{profile_cond}
	 \lim_{r \rightarrow \infty} \Delta(r) = \Delta_{\infty}  \quad \text{and}\quad \lim_{r \rightarrow 0} \Delta(r) = 0\, .
\end{equation}
In the following, we analyze the pairing-profiles
\begin{align}
  \Delta_{\text{a}}(r) &=  \Delta_{\infty}\Theta(r-R)\, , \label{eq:delta_a}\\
  \Delta_{\text{b}}(r) &= \Delta_{\infty}\begin{cases} r/R\, , & r<R \, , \\ 1\, , & r > R\, , \end{cases} \label{eq:delta_b}\\
  \Delta_{\text{c}}(r) &= \Delta_{\infty}\tanh(r/R)\, \label{eq:delta_c},
\end{align}
in details, see Fig.~\ref{fig:prof_sketches}. Different from the
discussion of the in-gap spectrum in superconductors, for heterostructures, we
keep the radius $R$ of the vortex as a free parameter independent of the
coherence length $\xi = \hbar v_D /\Delta_{\infty}$ in the TI. The term
`natural' vortex will refer to the fact that the radius of the vortex
coincides with the coherence length \footnote{We call the relation $R=\xi$
natural as this allows for comparison with the results of
Ref.~\cite{CAROLI-1964} under the identification $\mu \mapsto E_F$ and $v_D
\mapsto v_F$ of the parameters in the semiconductor with the ones in the
superconductor.}. We use $\Delta_\text{a}$ with the sharp pairing profile as a
model for an etched hole in the superconductor. The profile $\Delta_\text{c}$ on the other hand corresponds
to a magnetic vortex that is trapped inside the SC with $R = \xi_\text{SC}$ the coherence length of the superconductor. 
We show in Fig.~\ref{fig:spec_comp_TI_ab} that $\Delta_\text{b}$ approximates the latter well. As the ramp profile $\Delta_{\text{b}}$ allows for explicit analytical expressions, we will concentrate on $\Delta_\text{a}$ and $\Delta_\text{b}$ in the following.


\section{Excitation gap of the Majorana mode\label{sec:energie_gaps_TI}}

\subsection{Results and comparison to numerics}

Due to the presence of the superconductor, gating capabilities of the TI
surface are limited and we therefore focus on the regime where the chemical
potential is larger than the Thouless energy, $\mu \gg \hbar v_D/R$. 
We first present the results and move the derivation to the next subsection.  
We find that the energies of the in-gap Andreev states are given by $E_m = m\, \delta$, $m\in\mathbb{Z}$, with the level spacing
\begin{align}\label{E_gap_TI}
	\delta &= \delta_{\Delta} + \delta _{A}\notag \\
	&=\frac{\hbar v_D}{\mu} \left\langle \frac{\Delta(r)}{r} \right\rangle_w + \frac{{\pi \hbar}^2v_D^2 }{\Phi_0 \mu} \left\langle \frac{ A(r)}{r} \right\rangle_w \, ;
\end{align}
here and below, we defined the average
\begin{align}\label{averging}
  \langle f(r) \rangle_w &= \frac{\int_0^{\infty}\! dr\, f(r) w(r )}
  { \int_0^{\infty} \!dr\, w(r)}  \, ,
\end{align}
with the weighting function $w(r) = e^{- \frac{2}{\xi} \int_0^{r} \Delta(r')/
\Delta_{\infty}dr'}$, see Fig.~\ref{fig:prof_sketches}(b). The weighting
function is close to one inside the vortex core and decays for larger
radii. Note that the weighting function  for $\Delta_{\text{b}}$ approximates
the one for $\Delta_{\text{c}}$ rather well. Note that the results (valid in the limit of a small Fermi wavelength) coincide with the findings of Ref.~\citep{deng2020bound} derived by a different method. 

Given a paring profile $\Delta(r)$ and magnetic field strength $B(r)$, Eq.~\eqref{E_gap_TI} immediately yields the in-gap spectrum. As such, it constitutes an important guidance for on-going experimental efforts. Note that in the self-consistent description of superconductors these two profiles are coupled and require the solution of, \emph{e.g.}, the Ginzburg-Landau equation for a specific setup \cite{GL_treatments2000}. 
The contribution of the magnetic term $\delta_A$ depends on the type of superconductor and the radius $R$. In order to obtain a result, we need the magnetic field $B(r)$ in the region of suppressed $\Delta$, where the weighting function $w(r)$ is largest. 
A vortex traps the flux $\Phi_0$ in an area of roughly $\pi (R+\lambda)^2$ centered at the normal core with $\lambda$ the London penetration depth. For type-II superconductors $(\xi \ll \lambda)$ , we assume a homogeneous magnetic field of size $B(r)= \Phi_0/\pi (R+\lambda)^2$ throughout the region of suppressed $\Delta$ and find the correction
\begin{equation}\label{E_mag}
	\delta_{A}\approx\frac{\hbar^2 v_D^2}{2 \mu(R+\lambda)^2}\, .
\end{equation}
We have to distinguish two cases. For $\lambda \gg R\gtrsim \xi$, the magnetic correction is negligible \cite{CAROLI-1964}. On the other hand, for $\lambda \ll R$, the magnitude of $\delta_A$ scales as $\delta_{A} \simeq \hbar^2 v_D^2/2 \mu R^2\ll \Delta_{\infty}$. Therefore, this scaling self-consistently satisfies the condition for the perturbation theory presented in the subsequent section.
For a type-I superconductor ($\lambda \ll \xi \lesssim R$) the correction needs to be evaluated separately, since the full flux quantum penetrates the region occupied by the low-energy states. Considering a strong type-I superconductor with the homogeneous field $B(r)= \Phi_0/\pi R^2$ for $r<R$ and zero outside, we obtain
\begin{equation}\label{eq:E_mag_type1}
	\delta_{A}\approx \frac{\hbar^2 v_D^2}{4 \mu R^2}\frac{4R+\xi}{2 R + \xi} \stackrel{(R \gg \xi)}{\longrightarrow} \frac{\hbar^2 v_D^2}{2 \mu R^2 }\, .
\end{equation}
Again, the result self-consistently satisfies the perturbation
theory and we observe that Eq.~\eqref{E_mag} also holds for type-I, where we have $R\gg\lambda$. 
We expect this expression to work rather well for the profile $\Delta_{\text{a}}$, where no screening currents are present in the relevant region $r \lesssim R$ and give a reasonable estimate for the arbitrary gap profiles. 
In the following, we will be focusing on the second term $\delta_\Delta$.

Evaluating the contribution of the pairing term for the profiles $\Delta_\text{a}$ and $\Delta_\text{b}$ yields
\begin{align}
  \delta _{\Delta_\text{a}} &\approx\frac{\hbar^2 v_D^2}{4 \mu R^2}\frac{4R-\xi}{2 R + \xi} \stackrel{(R \gg \xi)}{\longrightarrow} \frac{\hbar^2 v_D^2}{2 \mu R^2 }\, ,\label{E_box_TI_gap} \\
	\delta _{\Delta_\text{b}} &\approx\frac{\hbar v_D \Delta_{\infty}}{\mu R}\, ,\label{E_ramp_TI_gap}
\end{align}
which is valid for  $\mu \gg \Delta_{\infty} \gtrsim \hbar v_D /R$. Note that for a sharp step, the level spacing that determines the minigap agrees in scaling with quantization modes in a box and decreases faster than for the ramp profile. Because of this, it is preferable not to etch the superconductor but to have magnetic flux line trapped inside the SC \footnote{The finite size quantized energy scale coincides with the Thouless energy $E_{\text{th}}=\hbar v_D / R$ for the clean system. We expect these formulas to generalize to disordered topological insulators by replacing the Thouless energy with its definition in the diffusive case.}.
Irrespective of the gap profile, gating capabilities are needed to lower $\mu$ and ensure that the energy spacing $\delta$ is experimentally resolvable. This agrees with the findings of Refs.~\citep{Das_Sarma2010} and \citep{deng2020bound} for natural vortices.

\begin{figure}[bt]
  \centering
\includegraphics[width = \linewidth]{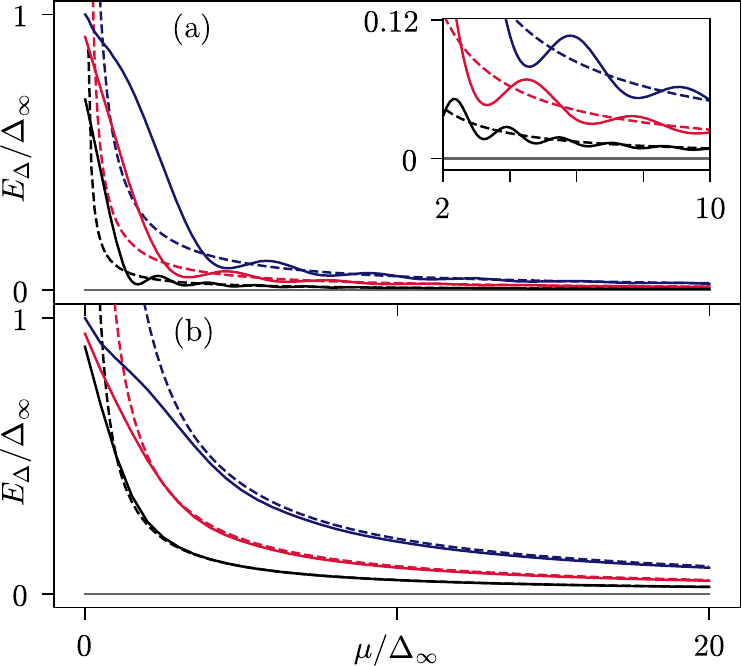}
\caption{Panels (a) and (b) compare the numerically determined spectra of the full Hamiltonian $H_R$ dependent on the chemical potential $\mu$ (solid lines)  with the analytic approximation $E_m = m\, \delta_{x}$ of Eqs.~\eqref{E_box_TI_gap} and \eqref{E_ramp_TI_gap}. Panel ($x$), $x\in\{a,b\}$, shows the energy levels for a pairing profile $\Delta_{x}(r)$. The gray, red, blue lines correspond to $m \in \{ 0,1,2\} $ for a `natural' radius $R=\xi$.  The black line depicts the result for $R=2\xi$ and $m=1$. Note that the numerically obtained spectra agree well with the predictions in (b). For the step potential in (a), additional oscillations appear due to the sharp features in $\Delta_\text{a}(r)$.} 
\label{fig:spec_full}
\end{figure}

In Fig.~\ref{fig:spec_full}, we compare the numerical solution of $H_R$ to the approximate expressions from Eq.~\eqref{E_gap_TI}. A multidimensional shooting method is employed to calculate the exact solutions throughout this paper; for details see App.~\ref{app:shooting}.
In the case of $\Delta_{\text{b}}$, the analytic solution agrees with the numerical solution for different values of $m$ and $R$. Using the step profile $\Delta_{\text{a}}$, however, the approximation for $E_m$ only captures the average value of the energy, omitting a significant oscillating contribution. The latter can be attributed to interference effects of the radially oscillating wavefunctions within the extended core region with $\Delta=0$. Note that breaking the rotational symmetry or introducing disorder in the core region should destroy these interference effects (cf. Sec.~\ref{sec:osc_both}). 

We can evaluate Eq.~\eqref{E_gap_TI} for $R=\xi$, \textit{i.e.}: natural vortices. We find $\delta_{\Delta_x} = \beta_x \Delta^2_\infty/\mu$ with $\beta_a \approx 0.24$, $\beta_b \approx 0.95$, $\beta_c \approx 0.85$ which agrees with the  calculations of Ref.~\cite{deng2020bound}, where they evaluated the case b.

\subsection{Derivation\label{sec:dom_mu_TI}}

In this section, we derive the expression Eq.~\eqref{E_gap_TI} which was used in the last section to discuss our main findings. Different from previous semi-classical methods \cite{CAROLI-1964,deng2020bound}, we employ standard perturbation theory for this task. We split the Hamiltonian
\begin{equation}
  H_R = H^{(0)} + H_{\Delta}^{(1)}\, ,
\end{equation}
into the unperturbed part $H^{(0)}$ with the perturbation
$H_{\Delta}^{(1)} = m  (2\hbar v_D\Delta(r)/\mu r)\mathop{\rm diag}(0,1,1,0 ) $. The reason for this choice is that we can find an exact zero mode for $H^{(0)}$, see below. Note that the full radial Hamiltonian is given by $H_R +H_A$ such that $H_{A}^{(1)}=H_A$ also acts as a perturbative term.

We find that the unperturbed Hamiltonian $H^{(0)}$ has a solution  
\begin{equation}\label{eq:zm}
  {F}_{m}(r) = \sqrt{ r \,w(r)} \begin{pmatrix} J_{m-1}(\tfrac{ \mu r}{\hbar v_D}) \\J_{m}(\tfrac{\mu r}{\hbar v_D}) \\-J_{m}(\tfrac{\mu r}{\hbar v_D}) \\J_{m+1}(\tfrac{\mu r}{\hbar v_D})
	\end{pmatrix}\, ,
\end{equation}
to the energy $E=0$ for \emph{each} angular momentum $m\in \mathbb{Z}$; where $J_m$ is the $m$-the Bessel function. Thus, the unperturbed Hamiltonian $H^{(0)}$ features a zero mode in each angular momentum sector. 

The particle-hole symmetry requires that under the perturbation the state at $m=0$ stays at $E=0$ while the condition for the rest of the spectrum simply demands $E_m = -E_{-m}$.
To illustrate the existence of the exact zero mode at $m=0$, we solve the full Hamiltonian \eqref{eq:matrix_ham_bdg} for a strong type-I SC in App.~\ref{app:zero_mode}. This mode at $m=0$ and $E=0$ is a MZM.

In order for the perturbation theory to work, we need to make sure that the perturbative terms $H^{(1)}$ are small compared to the first excited state of $H^{(0)}$. We have numerically checked that the first excited state of $H^{(0)}$ is at energy $\Delta_{\infty}$ for $\mu \gg \hbar v_D / R$ irrespective of the pairing profile for radii of the order of the coherence length. To obtain the approximate spectrum for $H_R+H_A$, we determine the energy in first order as
\begin{align}\label{e_correction_general}
  E_m &=  \frac{ \int_0^{\infty}\!dr \,{F}^{\dagger}_{m}(r) (H_{\Delta}^{(1)} + H_{A}^{(1)}){F}_{m}(r) }{\int_0^{\infty} \!dr\, {F}^{\dagger}_{m}(r) {F}_{m}(r) } \,.
\end{align}
Before investigating the integrals explicitly, we give an estimate of the order of magnitude for $H_{\Delta}^{(1)}$. Due to the exponential term in the weight function $w(r)$, the probability distribution $|{F}_{m}(r)|^2$ has the highest weight for $r<R$. In contrast, the gap profile in $H^{(1)}_{\Delta}$ leads to dominant contributions for $r>R$. Therefore, we obtain an estimate of the energy by evaluating $H_{\Delta}^{(1)}$ at $r=R$, which yields
\begin{equation}\label{scale_estimate_TI}
	H^{(1)}_{\Delta} \simeq m\frac{\hbar v_D \Delta_{\infty}}{\mu R} = m\frac{\xi \Delta_{\infty}^2}{\mu R}\, .
\end{equation}
This reproduces the scaling from Refs.~\cite{CAROLI-1964,deng2020bound} for the case of a natural vortex with $R=\xi$.

The explicit form of Eq.~\eqref{e_correction_general} is given by
\begin{align}\label{e_explicit}
  E_m \!=\! \frac{4m \hbar v_D \Bigl\langle\bigl[ \Delta(r) + \tfrac{\pi\hbar v_D  A(r)}{\Phi_0} \bigr]  J_{m}^2(\tfrac{\mu r}{\hbar v_D}) \Bigr\rangle_w}{ \mu \Bigl\langle r \,[J_{m-1}^2(\tfrac{\mu r}{\hbar v_D})+2J_{m}^2(\tfrac{\mu r}{\hbar v_D})+J_{m+1}^2(\tfrac{\mu r}{\hbar v_D})] \Bigr\rangle_w}\,.
\end{align}
If  we assume that $\Delta(r)$ and $A(r)$ are slowly varying on the scale
$\mu/\hbar v_D$, we can replace $J_m(x)^2$ with $1/\pi x$ by averaging
over the fast oscillating component of the Bessel function, see
App.~\ref{app:Bessel_osz}.   As a result, we obtain
$E_m \approx m\, \delta$ with the
minigap $\delta$ given by Eq.~\eqref{E_gap_TI} above.

\section{The electron gas heterostructure \label{sec:de-Gen-syst}}

For comparison, we consider in this section the case of a conventional 2DEG such  as GaAs with a quadratic spectrum that is proximity coupled to a SC. In particular, this allows to connect to earlier results that have been obtained for in-gap spectra of vortex lines in pure type-II superconductors \cite{CAROLI-1964}. The Hamiltonian is given as $H' = \int\! d^2r\, {\Psi}^{\dagger}(\bm{r}) H'_{\text{BdG}}(\bm{r}) {\Psi}(\bm{r})$ where we choose ${\Psi}(\bm{r}) = ( \psi_{\uparrow}(\bm{r}), \psi^\dag_{\downarrow}(\bm{r}) )^T $. The Bogoliubov Hamiltonian is given by ($m_*$ is the effective mass)
\begin{equation}
  H'_{\text{BdG}} = \begin{pmatrix}
    \tfrac{1}{2m_{*}} \left( \bm{p}+ \tfrac{e}{c} \bm{A} \right)^2 -\mu  & \Delta(\bm r) \\
    \Delta^*(\bm r) & \mu - \tfrac{1}{2m_{*}} \left( \bm{p}- \tfrac{e}{c} \bm{A} \right)^2 
\end{pmatrix}
    \, 
\end{equation}
where we have neglected the Zeeman energy.
The calculation proceeds analogously to Sec.~\ref{sec:phys_model} where, here, we employ  the ansatz
\begin{equation}\label{trafo_2DEG}
  {\Psi}(\bm{r}) = e^{i (m -\frac{1}{2}\sigma_z )\varphi} \frac{G(r)}{\sqrt{r}}\,.
\end{equation}
Uniqueness of the wave function now requires $m\in \mathbb{Z}+\tfrac12$. The application of Eq.~\eqref{trafo_2DEG} leads to the real Hamiltonian for the radial motion $H'=H'_R + H'_A$, which we split in the magnetic term 
\begin{equation}
	H'_A=\frac{\pi \hbar^2 A(r)}{2 m_* \Phi_0 r}\left(2m -\sigma_z + \frac{\pi r A(r)}{\Phi_0}\sigma_z \right)
\end{equation}
and the remaining parts
\begin{align}
  H_R' =   -\frac{\hbar^2}{2 m_*}\left( \sigma_z \partial^2_r -\frac{ m^2 \sigma_z - m }{r^2}  \right)+\Delta(r) \sigma_x - \mu \sigma_z  \, .\label{radial_2DEG}
\end{align}
Note that an important difference to Sec.~\ref{sec:phys_model} is the fact that in this case $m$ is half-integer. Because of this, all the states will be gapped and no zero modes remain.

\section{Electron gas energy gap\label{sec:de-Gen-spec}}

\subsection{Results and comparison to numerics}

As above, we first present our main findings. The in-gap spectrum is approximately given by $E'_m = m \,\delta'$.  For $\mu\gg \hbar^2 / m_*R^2$, the level spacing assumes the form
\begin{align}\label{E_simple_large_mu_2DEG}
	\delta' &= \frac{\hbar}{\sqrt{2m_* \mu}} \left\langle \frac{\Delta(r)}{r} \right\rangle_w + \frac{\pi \hbar^2}{\Phi_0 m_*} \left\langle \frac{A(r)}{r} \right\rangle_w \, .
\end{align}
Note that comparing \eqref{E_simple_large_mu_2DEG} to \eqref{E_gap_TI}, we see
that with the correspondence $v_F =\sqrt{2\mu/m_*} \leftrightarrow v_D$, we
find $\delta' =\tfrac{1}{2}\delta$. Rewriting the first summand $\delta'_{\Delta}$ in Eq.~\eqref{E_simple_large_mu_2DEG} in terms of $k_F$, it coincides with the formula derived in Refs.~\cite{CAROLI-1964, deng2020bound}.

\begin{figure}[tb]
  \centering
\includegraphics[width = \linewidth]{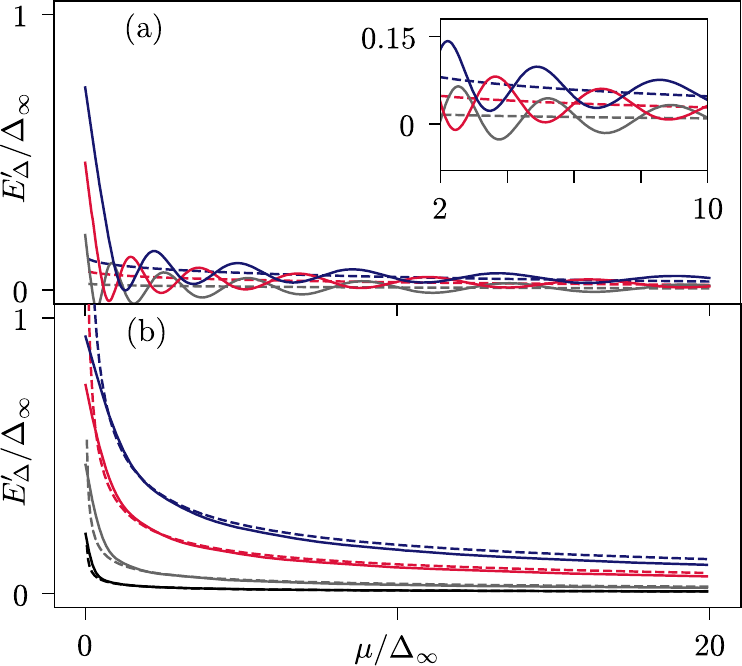}
\caption{Comparison of the numerically determined spectra of the full Hamiltonian $H_R'$ as a function of the chemical potential (solid lines) $\mu$ with the analytical prediction $ E'_m= m\,\delta'_x = \tfrac12 m\, \delta_x$ of Eqs.~\eqref{E_box_TI_gap} and \eqref{E_ramp_TI_gap} with the replacement $v_D \mapsto v_F$. Panel ($x$), $x\in\{a,b\}$, show the energy levels for a pairing profile $\Delta_{x}(r)$. The gray, red, blue lines correspond to the choices $m \in \{\tfrac12,\tfrac32,\tfrac52 \} $ with  $\hbar/m_{*} R^2=0.1 \Delta_{\infty}$. The black line depicts $m=\tfrac12$ for $\hbar/m_{*} R^2= 0.01\Delta_{\infty}$. Note that the numerically determined values agree well with the analytical predictions for $\Delta_{\text{b}}$. In panel $(a)$ the additional oscillations are due to the sharp step potential, see Sec.~\ref{sec:osc_both}. }
\label{fig:spec_full_dG}
\end{figure}

As for the TI heterostructure, we find that the minigap decreases with increasing $R$. Analogous to section~\ref{sec:energie_gaps_TI}, approximating the magnetic field to be homogeneous yields
\begin{equation}
  \delta'_A = \frac{\hbar^2}{2 m_*(R+\lambda)^2} \, .
\end{equation}
The strength of this term again depends strongly on the type of superconductor. In the following, we focus on the correction due to the pairing potential. The comparison between the $R\gtrsim \xi$ approximation to $\delta'_{\Delta}$ in Eq.~\eqref{E_simple_large_mu_2DEG} and the numerical solution to Eq.~\eqref{radial_2DEG} is displayed in Fig.~\ref{fig:spec_full_dG}. 

As for the TI, we evaluate the expression explicitly for natural vortices with $R=\xi$ and find $\delta'_{\Delta_x} = \beta'_x \Delta^2_\infty/\mu$ with $\beta'_\text{a} \approx 0.12$, $\beta'_\text{b} \approx 0.47$, $\beta'_\text{c} \approx 0.43$.
Thus, the results agree with the $\Delta_{\text{SC}}^2/E_F$ of Ref.~\cite{CAROLI-1964} and reproduce the factor $\tfrac12$ obtained in Ref.~\citep{deng2020bound} for case b.

\subsection{Derivation\label{sec:dom_mu_2DEG}}
Analogously to Sec.~\ref{sec:dom_mu_TI}, we want to set up a perturbative treatment of $H_R'$ valid in the limit $\mu\gg \hbar^2 / m_*R^2$. We write $H'_R= H'^{(0)} + H_{\Delta}'^{(1)}$ with 
\begin{align}\label{diag_for_2DEG}
  H_{\Delta}'^{(1)} &= m \frac{\hbar }{\sqrt{2 m_* \mu}} \frac{\Delta(r)}{r}+\left(\frac{ \hbar\partial_r\Delta(r)}{\sqrt{8m_* \mu}}  - \frac{\Delta^2(r)}{4 \mu}\right)\sigma_z\,.
\end{align}
As before, we have a zero mode $G_m(r)$ of $H'^{(0)}$ in each sector $m$ given by
\begin{equation}
  {G}_{m}(r) =  \sqrt{r\, w(r)} \begin{pmatrix} J_{m-\frac{1}{2}}(\tfrac{\sqrt{2m_* \mu}\, r}{\hbar}) \\J_{m+\frac{1}{2}}(\tfrac{\sqrt{2 m_* \mu}\, r}{\hbar })
	\end{pmatrix}\, .
\end{equation}
The spectra of $H'^{(0)}$ are gapped by $\Delta_{\infty}$ for $  \mu\gg \hbar^2 / m_*R^2$, irrespective of the pairing profile for radii of the order of the coherence length. For slowly changing $\Delta(r)$ and $A(r)$, we obtain 
$E'_{m} \approx m \,\delta'$ with $\delta'$ given by Eq.~\eqref{E_simple_large_mu_2DEG} analogous to section~\ref{sec:dom_mu_TI} (details see App.~\ref{app:Bessel_osz_2DEG}).
Notably, the correction terms proportional to $\sigma_z$ drop out. 

\section{Oscillation corrections for step pairing potential\label{sec:osc_both}}

\begin{figure}
  \centering
\includegraphics[width = \linewidth]{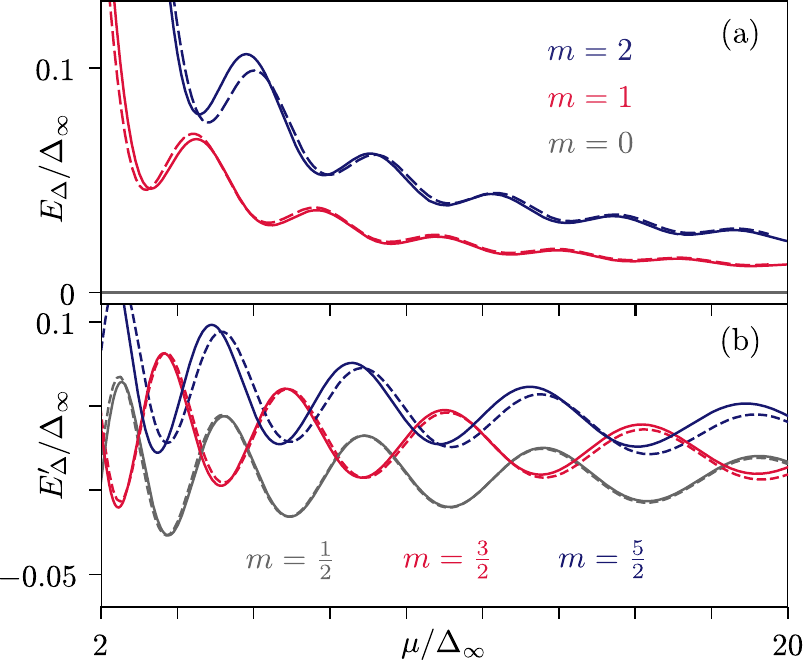}
\caption{Zoom into the oscillatory behavior of the in-gap spectrum for the step pairing potential $\Delta_\text{a}(r)$ for different angular momenta $m$. Panel (a) [(b)] shows the results for the topological insulator [two-dimensional electron gas]. The analytical predictions of Eqs.~\eqref{osc_corr_TI}, valid for $\mu \gg \hbar v_D/R$, and \eqref{osc_corr_2DEG}, valid for $\mu \gg \hbar^2/m_* R^2$, are shown as dashed lines. The solid lines show the full solutions using the Hamiltonian $H_R$ and $H_R'$, respectively. We have chosen  $R=\xi$ in (a) and $\hbar/m_{*} R^2=0.1\Delta_{\infty}$ in (b). The oscillations are captured better for smaller $m$ and their amplitude decreases for large $\mu$. Note that while the oscillations decrease faster than the mean in (a), for panel (b) they both scale as $\mu^{-1/2}$ such that we observe many energy level crossings as well as sign changes in this case.}
\label{fig:spec_full_osc}
\end{figure}

In this section, we would like to understand the oscillations of the in-gap spectrum that appear for the pairing profile $\Delta_\text{a}$, see insets of Figs.~\ref{fig:spec_full} and \ref{fig:spec_full_dG}. In particular, they arise due to the fact that $\Delta_\text{a}(r)$ has a sharp feature close to $r=R$ invalidating the replacement $J_{\alpha}^2(x) \mapsto  1 /\pi x$. A better approximation is given by $J_{\alpha}^2(x) \mapsto  (2 /\pi x) \cos^2\left( x - \tfrac{\pi}{2}\alpha - \tfrac{\pi}{4}\right) $
in Eq.~\eqref{e_explicit} and \eqref{e_explicit_2DEG}. In case of Eq.~\eqref{e_explicit}, the incorporation of the oscillation is straight forward and we obtain ($\mu \gg \Delta_{\infty} \gtrsim \hbar v_D/R$)
\begin{align}\label{osc_corr_TI}
  E_{m} &=  m \delta_{\Delta_\text{a}} + (-1)^{m} m \frac{\hbar^2 v_D^2\cos(2\mu R/\hbar v_D)}{\mu^2R(2R + \xi)}\Delta_{\infty} \, .
\end{align}
This result matches well with the numerical solution of $H_R$, cf. Fig.~\ref{fig:spec_full_osc}(a). Note that the oscillations are suppressed by a factor of $\hbar v_D/R\mu \ll 1$. 
For the 2DEG, the oscillating correction is analogously given by ($\mu \gg \Delta_{\infty} \gtrsim \hbar^2 / m_*R^2$)
\begin{align}\label{osc_corr_2DEG}
  E'_{m} &= m \delta'_{\Delta_\text{a}} + (-1)^{m-1/2}\frac{\hbar\sin(\sqrt{8m_* \mu}\, R/\hbar )}{\sqrt{2m_*\mu}(2R + \xi)} \Delta_{\infty}\, .
\end{align}
The perturbative result is compared to the exact solution of the full problem in Eq.~\eqref{radial_2DEG} in Fig.~\ref{fig:spec_full_osc}(b). Note that in this case, the oscillation amplitudes are of the same magnitude as the leading term for $m=\frac12$. Because of this, the oscillations for small $m$ lead to crossings at $E=0$.

While our theory accurately predicts the oscillating behavior of the in-gap energies for a sharp gap profile, we do not expect these oscillations to be relevant for the experiments. The reasons is that they are easily washed out by either disorder or imperfect radial symmetry. The latter is most relevant for the 2DEG as many levels (of different angular momentum) cross at low energies, see Fig.~\ref{fig:spec_full_osc}(b).

\section{Conclusion \label{sec:concl}}

In this work, we considered in-gap states of vortices in clean
heterostructures. Using Rayleigh-Schr\"odinger perturbation theory, we rederived the concise formula 
Eq.~\eqref{E_gap_TI} of Ref.~\cite{deng2020bound} to predict the energy gap above the Majorana zero mode in
vortices in proximitized topological insulators. With this formula, the effects of the magnetic field profile on the energy difference of the in-gap states is predicted, in addition to its pairing profile dependence. We have compared different design principle given by different gap profiles. We have found that it is unfavorable to etch the superconductor if the radius is larger than the coherence length $\xi$. We have additionally discussed the orbital effects due to the magnetic field for type-I superconductors. We have found that the magnetic field contribution is of the same order as the pairing gap contribution. Our results can guide experimental efforts aiming at increasing the energy
separation of the Majorana zero mode to finite energy excitations. We have checked the approximations employed by comparing our results with numerics.

\begin{acknowledgments}
We acknowledge fruitful discussions with Jakob Schluck. This work was funded by the Deutsche Forschungsgemeinschaft (DFG, German Research Foundation) under Germany's Excellence Strategy -- Cluster of Excellence Matter and Light for Quantum Computing (ML4Q) EXC 2004/1 -- 390534769.
\end{acknowledgments}

\appendix

\section{Multidimensional shooting method\label{app:shooting}}

In this section, we discuss the multidimensional shooting method that we used to numerically solve the systems of radial equations \eqref{radial_TI} and \eqref{radial_2DEG}. The idea is to match the boundary conditions at $r=0$, propagate the solution in a shooting step to a final position $R' \gg R$ and see for which energies the wave functions can satisfy the boundary condition at the final point.

\begin{figure}
  \centering
\includegraphics[width = \linewidth]{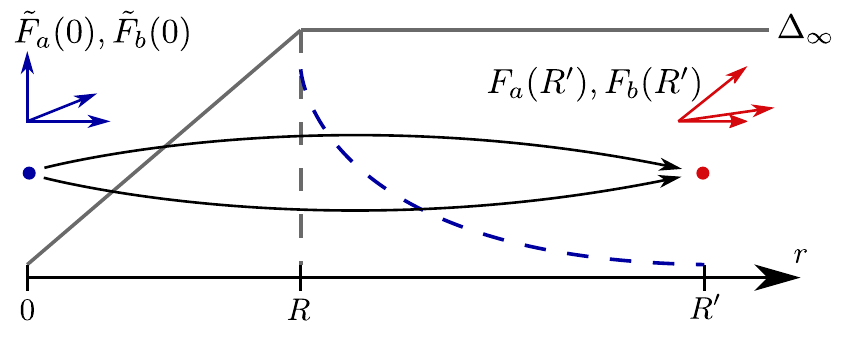}
\caption{Schematic of the multidimensional shooting method. Iterate through the sub-gap energies and propagate the asymptotic solutions $\tilde{F}_a$ and $\tilde{F}_b$ for $r \to 0$ using a standard ODE solver to the solutions $F_a$ and $F_b$ at $R'\gg R$. The boundary value problem is solved if we can construct an approximate zero state using a linear combination of $F_a$ and $F_b$. The latter has to hold in order to have a normalizable solution for $r \rightarrow \infty$.}
\label{fig:shooting}
\end{figure}

For small $r$, we can neglect the pairing and obtain the pair of solutions [$x_\pm = (\mu \pm E) r/\hbar v_D$]
\begin{equation}
  \frac{\tilde F_a(r)}{r^{1/2}} \!=\! \begin{pmatrix} 
    J_{m-1}(x_+)\\
       J_{m}(x_+)\\
	0 \\
	0
        \end{pmatrix}, \quad  \frac{\tilde F_b(r)}{r^{1/2}} \!=\!  \begin{pmatrix}
	0\\
	0\\
         -J_{m}(x_-)\\
      J_{m+1}(x_-)	
    \end{pmatrix}  
\end{equation}
of Eq.~\eqref{radial_TI} (with $\Delta =0$). Using a standard ODE solver, we find solutions $F_a$ and $F_b$ of the differential equation  \eqref{radial_TI} (for finite $\Delta$ and $r\in [0,R']$)  that coincide with $\tilde F_a$ and $\tilde F_b$ for $r\to0$, see Fig.~\ref{fig:shooting}. The task then is to find the energy $E$ for which there exists a linear superposition $F(r) = c_a F_a(r) + c_b F_b(r)$ where $F(r \to \infty)_j \to 0$ (for each component $j$).
For this it is necessary that the $2\times2$ matrix 
\begin{equation}
  M_{12} = \begin{pmatrix}
    F_a(R')_1 & F_a(R')_2\\
    F_b(R')_1 & F_b(R')_2
  \end{pmatrix}
\end{equation}
has a vanishing determinant for $R' \to \infty$. For our numerical results, we have evaluated $\det (M_{12})$ for $R' =R + 20 \xi$ as a function of $E$ and found its roots. We have checked that the results are the same choosing $M_{13}, M_{24}$ and $M_{34}$. We cannot choose the orbitals that are related by particle hole symmetry ($1 \leftrightarrow 4$, $2 \leftrightarrow 3$).

The results of Eq.~\eqref{radial_2DEG} for the 2DEG are obtained similarly by replacing $\tilde F_a$ and $\tilde F_b$ with [$x_\pm = \sqrt{2 m_* (\mu \pm E)} r/\hbar$]
\begin{equation}
  \frac{\tilde G_a(r)}{r^{1/2}} \!=\!\begin{pmatrix} J_{m-\frac{1}{2}}(x_+) \\0   \end{pmatrix},\;   \frac{\tilde G_b(r)}{r^{1/2}} \!=\!\begin{pmatrix} 0\\J_{m+\frac{1}{2}}(x_-)
	\end{pmatrix}\, .
\end{equation}

\section{Zero mode\label{app:zero_mode}}
In this section, we discuss the zero mode of the full Hamiltonian Eq.~\eqref{eq:matrix_ham_bdg} for pairing profile $\Delta_{\text{a}}$ and strong type-I SC for which 
the magnetic field $B(r)= \Phi_0/\pi R^2$ is homogeneous within the region of  suppressed pairing profile and zero outside. In the chosen gauge, the respective vector field follows to
\begin{equation}
	A(r) = \begin{cases} \frac{\Phi_0}{2 \pi R^2} r, & r\leq R \\
	\frac{\Phi_0}{2 \pi r}, & r > R \, .
	\end{cases}
\end{equation}
We solve the Schrödinger equation for $E=0$ in the inside ($r < R$) and outside $(r > R)$ separately and match the solutions at $r=R$. In the inside, the wave function results in [with $x= \mu R/\hbar v_D $]
\begin{equation}\label{eq:zero_mode_inside}
	\Psi_{<}(r) = N e^{-\tfrac{r^2}{4 R^2}} \begin{pmatrix}
          i e^{-i\varphi} \frac{r}{x R} \mathop{\rm M}\Bigl(1-\tfrac12 x^2,2,\frac{r^2}{2 R^2}\Bigr) \\
          \frac{1}{x^{2}} \mathop{\rm M}\Bigl(-\tfrac12 x^2,1,\frac{r^2}{2 R^2}\Bigr)  \\
          -\frac{i}{x^2} \mathop{\rm M} \Bigl(-\tfrac12 x^2,1,\frac{r^2}{2 R^2}\Bigr) \\
	 e^{i \varphi}\frac{r}{x R} \mathop{\rm M}  \Bigl(1-\tfrac12 x^2,2,\frac{r^2}{2 R^2}\Bigr)\, .
	\end{pmatrix}
\end{equation}
Here,
\begin{equation}
  \mathop{\rm M}(a,b,z) = \sum_{k=0}^{\infty} \frac{(a)_k}{(b)_k}\frac{z^k}{k!} \, ,
\end{equation}
is the Kummer function of the first kind 
with $(a)_k = a(a+1)(a+2)\cdots(a+k-1)$ and $N$ a normalization constant \citep{abramowitz_2008}. Note that the wave function \eqref{eq:zero_mode_inside} is the same as for a graphene quantum dot in an applied magnetic field \citep{Ihn_2008}. For  the outside region, we obtain 
\begin{equation} \label{eq:zero_mode_outside}
	\Psi_{>}(r) = c_1 N e^{ \tfrac{R-r}{\xi}} \begin{pmatrix}
	 -i e^{-i\varphi} \Big[J_{-\tfrac12}(\frac{r \mu}{\hbar v_D}) - c_2 J_{\tfrac12}(\frac{r \mu}{\hbar v_D})\Big] \\
	 \Big[c_2 J_{-\tfrac12}(\frac{r \mu}{\hbar v_D}) + J_{\tfrac12}(\frac{r \mu}{\hbar v_D})\Big])  \\
	-i  \Big[c_2 J_{-\tfrac12}(\frac{r \mu}{\hbar v_D}) + J_{\tfrac12}(\frac{r \mu}{\hbar v_D})\Big] \\
	- e^{i \varphi}\Big[J_{-\tfrac12}(\frac{r \mu}{\hbar v_D}) - c_2 J_{\tfrac12}(\frac{r \mu}{\hbar v_D})\Big] \, ,
	\end{pmatrix}
\end{equation}
with $c_1,c_2$ arbitrary constants. For the solution in Eq.~\eqref{eq:zm}, we see that the angular momentum quantum numbers of the orbitals at $m=0$ are given by $\pm \tfrac12 \pm \tfrac12$, where one summand stems from the Berry phase of the TI and the other is a result of the phase winding of the pairing potential. In contrast to this, the quantum numbers in Eq.~\eqref{eq:zero_mode_outside} follow to $\pm \tfrac12$ since the full superconducting flux quantum contained in the region $r< R$ cancels the effect of the phase winding. This is a manifestation of the Meissner effect arising due to the  self-consistency equations supplementing the mean-field equations.

\begin{figure}
  \centering
\includegraphics[width = \linewidth]{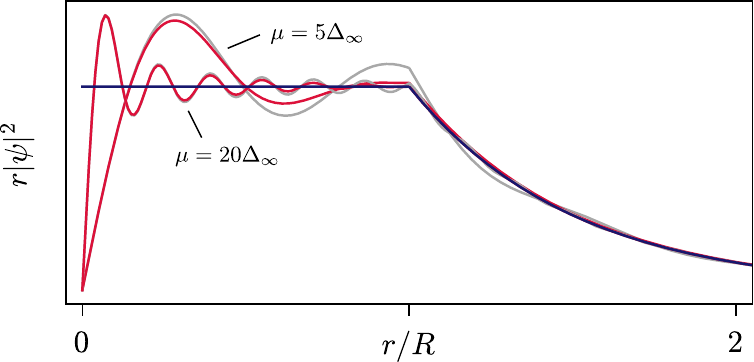}
\caption{This diagram displays the probability to find the Majorana zero mode at distance $r$ from the center of the vortex. This quantity is shown for both the exact zero energy wave function (red) from Eqs.~\eqref{eq:zero_mode_inside} and \eqref{eq:zero_mode_outside} as well as for the perturbative zero mode (gray) from Eq.~\eqref{eq:zm}. For large $\mu$, the probability to find the zero mode at $r$ tends for both cases towards a constant within the region of suppressed pairing potential and falls off according to $e^{-2(r-R)/\xi}$ outside (blue). For this plot $\xi=R$ was chosen.}
\label{fig:position_probability}
\end{figure}

\begin{figure}
  \centering
\includegraphics[width = \linewidth]{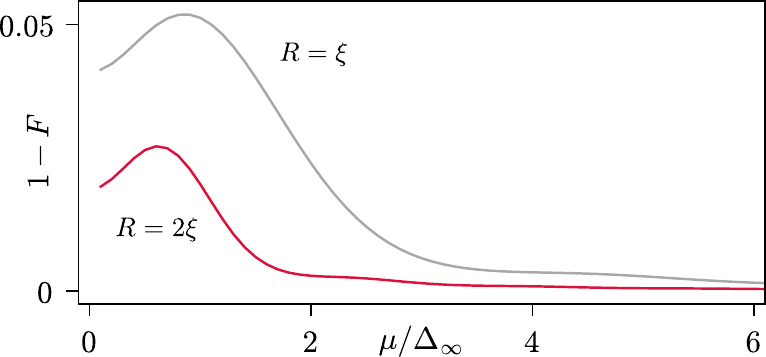}
\caption{Distance between the exact zero energy state from Eqs.~\eqref{eq:zero_mode_inside} and \eqref{eq:zero_mode_outside} and the perturbative zero mode from Eq.~\eqref{eq:zm} as a function of the chemical potential $\mu$. For the latter, the magnetic field is neglected. The fidelity $F$ is defined according to Eq.~\eqref{eq:fidelity}. In case of a natural vortex $R=\xi$ the distance between the states decreases with $\mu$ such that for $\mu \gg \Delta_{\infty}$ the effects of the magnetic field on the zero mode are negligible. This tendency only gets stronger upon increasing $R$.}
\label{fig:fidelity}
\end{figure}

Matching the wave functions at $r=R$ determines $c_1$ and $c_2$. In general these are straightforward to obtain but rather lengthy. In the limit $\mu \gg \hbar v_D /R$, we find the approximate expressions
\begin{align}
  c_1 \approx  \frac1{\sqrt{2}}\Bigl(\frac{\hbar v_D}{\mu R}\Bigr)^2 \qquad c_2 \approx \frac{2\hbar v_D}{3 \mu R} + 1\, .
\end{align} 
To visualize the exact zero mode, we plot $r |\Psi|^2$, the probability to find the mode at a distance $r$ from the vortex core, as a function of $r$ in Fig.~\ref{fig:position_probability}.
Both, the zero mode in Eqs.~\eqref{eq:zero_mode_inside} and \eqref{eq:zero_mode_outside} and the perturbative wave function \eqref{eq:zm}, where the magnetic field is not yet taken into account, converge for $\mu \gg \Delta_{\infty}$ to a constant probability for $r< R$, which then decays exponentially in the proximitized region. This means that irrespective of the specific magnetic field, the zero mode is equally likely found at any arbitrary distance $r< R$ for large chemical potential. 
Furthermore, we observe that as $\mu$ increases, the perturbative zero mode seems to approximate the exact zero mode more closely. To verify this, we use the fidelity 
\begin{equation}\label{eq:fidelity}
	F(\Psi_1,\Psi_2) = \Bigl| \int\! d^2r \,\Psi_1^{\dagger} \Psi_2  \Bigr|^2
\end{equation}
to measure the closeness of two wave functions $\Psi_1$ and $\Psi_2$. In Fig.~\ref{fig:fidelity} we show the distance $1-F$ between the state in Eqs.~\eqref{eq:zero_mode_inside} and \eqref{eq:zero_mode_outside} and the one from \eqref{eq:zm} as a function of the chemical potential $\mu$. For $\mu > \Delta_{\infty}$ both states converge and the magnetic field effects are thus negligible. This tendency is enhanced for larger $R$.

\section{TI integral solutions for $\mu \gg \hbar v_D / R$ \label{app:Bessel_osz}}

In this section, we discuss the approximations to the integral expression in Eq.~\eqref{e_explicit} and evaluate it explicitly for the pairing profile $\Delta_{\text{a}}(r)$ and $A(r)=0$. The first order corrections are non-zero only for $m>0$. In those cases, the contributions to the integral vanish for small $\mu r/ \hbar v_D \ll 1$ for all $\alpha \geq 0$ using Eq.~\eqref{profile_cond}. Thus, we approximate the Bessel function for large argument $\mu r/\hbar v_D \gg 1$ in agreement with the limitation of the perturbation theory that led to Eq.~\eqref{e_explicit}. Therefore, we replace
\begin{equation}\label{Bessel_osc_form_2DEG}
  J^2_\alpha(x) \simeq \frac{2}{\pi x} \cos^2(x-\tfrac\pi2  \alpha  - \tfrac\pi4 ) \,.
\end{equation}
With the proviso that $\Delta(r)$  does not change much on the scale $\hbar v_D/ \mu$, we can moreover replace $\cos^2(x)$ by its average over one period with the result
\begin{equation}\label{Bessel_ave_form_2DEG}
  J_{\alpha}^2(x) \simeq \frac{1}{\pi x} \, .
\end{equation}
Subsequently, the integral in Eq.~\eqref{e_explicit} is explicitly solved for
the step profile $\Delta_{\text{a}}(r) = \Delta_{\infty} \Theta(r-R)$. Inserting the averaged Bessel functions into Eq.~\eqref{e_explicit} results in
\begin{align}\label{ave_TI_Ea}
  E_{m} &= m\frac{\Delta_{\infty}^2}{\mu} \frac{2 e^{2X} \mathop{\rm E_1}(2X)}{2X+1} 	\approx m\frac{\Delta_{\infty}^2}{\mu}\frac{1}{4 X^2} \frac{4X-1}{2 X+1} \, ,
\end{align}
where we defined $X=R/\xi$ and approximated $X \gtrsim 1$, meaning $\Delta_{\infty} \gtrsim \hbar v_D / R$ in the second step. Additionally, we used the exponential integral
\begin{equation}
	\mathop{\rm E_1}(x) = \int_x^{\infty} \frac{e^{-t}}{t} dt\, .
\end{equation}
If we do not average the oscillations and insert Eq.~\eqref{Bessel_osc_form_2DEG}, we instead find 
\begin{align}
  E^\text{osc}_{m}&= E_{m} +(-1)^{m}m \frac{ \Delta_{\infty}^2}{\mu}\frac{2 e^{2X}}{2X+1}\text{Im}\bigl[ \mathop{\rm E_1}(\tfrac{2X(1 - i\mu) }{ \Delta_{\infty}}) \bigr] \notag \\
	&\approx E_{m} + (-1)^{m} m \frac{ \Delta_{\infty}^3}{\mu^2}\frac{ \cos(\tfrac{2X \mu }{ \Delta_{\infty}})}{X(2X+1)}\, ,
\end{align}
where we approximated $\mu\gg \Delta_{\infty} \gtrsim \hbar v_D /R$ in the second step. Besides the average energy contribution, there is an oscillatory term that is suppressed by an additional prefactor $\hbar v_D/R\mu$. Analogously, the corrections of the different pairing profiles can be calculated. Note that using the approximation Eq.~\eqref{Bessel_osc_form_2DEG} does not lead to significant oscillating corrections for the two remaining profiles $\Delta_{\text{b}}$ and $\Delta_{\text{c}}$ discussed here.

\section{2DEG integral solutions for $\mu\gg \hbar^2 / m_*R^2$ \label{app:Bessel_osz_2DEG}}

In this section, we analyze the first order correction for the spectrum in the 2DEG heterostructure and explicitly evaluate it for $\Delta_{\text{a}}$. Assuming $\mu\gg \hbar^2 / m_*R^2$ and setting $A(r)=0$, the correction for Eq.~\eqref{diag_for_2DEG} takes the integral expression  ($k_F = \sqrt{2 m_* \mu}/\hbar$)
\begin{widetext}
\begin{align}
  E'_{m} &= \frac{ \int_0^{\infty} d r~   w(r) \left[ J_{m+\frac{1}{2}}^2 (k_F r) \left(2 m\Delta(r)  -r \partial_r\Delta(r) + \frac{k_F r\Delta^2(r)}{2\mu}\right) +  J_{m-\frac{1}{2}}^2 (k_F r)\left(2 m\Delta(r) + r \partial_r\Delta(r) - \frac{k_F r \Delta^2(r)}{2\mu} \right)  \right]}{ 2\int_0^{\infty} d r\, k_Fr \, w(r) \left[ J_{m+\frac{1}{2}}^2(k_F r)+J_{m-\frac{1}{2}}^2(k_F r) \right]}\, . \label{e_explicit_2DEG}
\end{align}
\end{widetext}
Analogous to the previous section, the contribution to the integrals for $k_F r \ll 1$ vanishes and we approximate the Bessel functions for the dominant contribution with large arguments $x=k_F r\gg 1$. 

Note that the contributions of the second and third term in the perturbation \eqref{diag_for_2DEG} vanish, when we make use of Eq.~\eqref{Bessel_ave_form_2DEG}, but not for Eq.~\eqref{Bessel_osc_form_2DEG}. For the latter case, we analyze the relative strengths of the perturbation terms. 
Proceeding analogously to Eq.~\eqref{scale_estimate_TI}, with $\Delta(R)=\Delta_{\infty}$, we obtain as characteristic scales
\begin{align}
  E'_{m} & \simeq \frac{\Delta_{\infty}}{k_F R} + \frac{\partial_r\Delta(R)}{k_F} +\frac{ \Delta^2_{\infty}}{\mu} \notag \\
	&\simeq \frac{\xi \Delta_{\infty}^2}{\mu R} + \frac{\xi\Delta_{\infty} \partial_r\Delta(R)}{\mu} +\frac{ \Delta^2_{\infty}}{\mu} \,
\end{align}
where the first and third term resemble the Caroli-de-Gennes-Matricon scaling \cite{CAROLI-1964} for $R = \xi$. The slope of $\Delta(r)$ around position $R$ determines the relative strength of the remaining contribution. For $\Delta_{\text{b}}(r)$ and $\Delta_{\text{c}}(r)$, the average slope is sub-linear, such that the contribution can be neglected. For $\Delta_{\text{a}}(r) = \Delta_{\infty}\Theta(r-R)$ however, this term is expected to have a sizable effect due to the sharp feature. 
Subsequently, the integral in Eq.~\eqref{e_explicit_2DEG} is solved for $\Delta_{\text{a}}(r)$. Inserting Eq.~\eqref{Bessel_ave_form_2DEG} into Eq.~\eqref{e_explicit_2DEG} results in $E'_{m}=\tfrac12E_{m}$, 
but we need to consider $X=R/\xi$ being a $\mu$-dependent ratio. Note that in contrast to equation~\eqref{ave_TI_Ea}, $m$ takes half-integer values.
If we do not average the oscillations and insert Eq.~\eqref{Bessel_osc_form_2DEG}, we instead find 
\begin{align}
  E_{m}^{\prime\text{osc}}\approx E'_{m}+\frac{\cos(\pi m- 2 k_F R)}{k_F(2R + \xi)}\Delta_{\infty}\, .
\end{align}
under the assumptions $\mu \gg \Delta_{\infty} \gtrsim \hbar^2/ m_* R^2$. Notably, the amplitude of the oscillation matches the average value and is not suppressed. 

\bibliography{vort_spec}

\end{document}